# THEORETICAL AND EXPERIMENTAL MODELLING OF BUBBLE FORMATION WITH CONNECTED CAPILLARIES IN LIQUID COMPOSITE MOULDING PROCESSES

Yanneck Wielhorski, Amine Ben Abdelwahed, Laurent Bizet and Joël Bréard

*Laboratoire Ondes et Milieux Complexes, LOMC, UMR 6294 CNRS, 53 rue de Prony, BP 540, 76058 Le Havre, France*
*Corresponding author's e-mail: yanneck.wielhorski@univ-lehavre.fr*

**ABSTRACT**: The void prediction in LCM processes sparks off interest within the composite material industry because it is a significant issue to keep the expected mechanical properties. The liquid properties, the preform geometry and the flow conditions impact the quantity of void entrapped inside the final product. The complex geometry of the reinforcement due to the arrangement of the bundles and the fibres is a key point to understand and quantify this phenomenon. This paper deals with both simple model networks which can occur inside a fabric representing connected capillaries, so-called "Pore Doublet Model (PDM)". A first is considering two capillaries converging on a node (T-junction) and a second is representing two capillaries interconnected with a supplying principle. These configurations can affect locally the evolution of flow fronts. First, experiments of bubble formed in a T-junction device have been performed and studied. Then a theoretical approach was proposed to forecast microvoid and macrovoid formation, by taking into account a supplying principle and arranged Washburn equation in forced filling.

## INTRODUCTION

The well-known LCM processes refers to composite manufacturing techniques where a resin is injected through a fibrous preform. During this process, the void formation can occur and gas (air) bubbles are also entrapped (Fig. 1.*i*), decreasing therefore the mechanical properties of the final composite material. The origins of this phenomenon stem mainly, on a one hand, from the complex geometry of the reinforcement and, on the other hand, the wetting properties between liquid and solid. The fibrous preform is often represented with mainly two pore scales: a macroscale between the bundles and a microscale within the bundles and then the flows are governed differently according to the scale, either by the Laplace pressure inside the tow or by the viscous force between the tows. Hence, the bubble formation can be explained by the competition between the capillary and the viscous forces that are compared inside the capillary number *Ca*. Indeed, these both kinds of flow induce locally a difference between the front positions leading consequently to the bubble formation that can be summarized as follows: the intra-tow void, so-called microvoid, which is located inside the bundle and the inter-tow void, also called macrovoid, between two consecutive bundles. More precisely, the microvoids occur at high capillary number *Ca* because the viscous flow overcomes the capillary forces whereas the macrovoids are obtained at lower *Ca* because the capillary pressure dominates the viscous pressure.

To quantify the void and attempt to minimize its rate inside composite materials, many experimental studies have been carried out [1-4]. However, it is often difficult to visualize bubble formation and transport during the LCM process. Therefore, to determine the final saturation, numerical works [5], experimental and theoretical studies in porous Pore-Doublet Model [4, 6] are ways to get round this difficulty. The latter is a widely used modelling to understand the dynamic of immiscible fluids in porous media [7, 8].

In this paper, we attempt to underline the influence the wetting effect and geometrical configuration on capillary flow on the bubble formation by using two "Pore Doublet Models": a T-junction device (Fig. 1.*ii*) and two parallel capillaries first continuously interconnected over a variable distance then connected by nodes (Fig. 4).

First, to study the wetting influence on bubble shape and length, we performed experiments by using T-junctions devices [6, 9-11] based on the physical mechanism of cross-flowing streams. Moreover, this is widely used as well in microchemical engineering as in microbiology so that obtain, for instance, calibrated and controlled amounts of liquids (reagents, polymer, etc.). This kind of device allows to study interactions between two immiscible fluids (liquid/liquid or liquid/gas) by converging a dispersed phase (break-up stream) with a continuous phase (shear stream). Two main regimes can be distinguished: the squeezing and the dripping regimes respectively based on confined and unconfined breakup mechanisms. The squeezing regime, occurring at low *Ca*, in which the interfacial force overcomes the viscous shear stress and thus the bubble is usually quite longer than the channel width (slug bubbles), will be showed here.

Secondly, we emphasise the supplying principle arising from the bulk provided by the macrochannel, which plays the role of a tank, to the microchannel [4, 6, 12]. A theoretical model will be presented by taking into account supplying principle within interconnected capillaries.

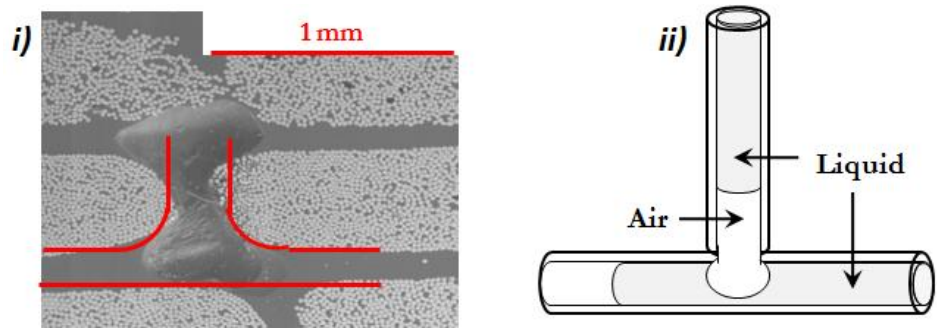

Fig. 1: *i)* Transversal view of a composite material with an entrapped air bubble and *ii)* its equivalent model sketch.

## BUBBLE FORMATION WITHIN PORE DOUBLET MODELS

### T-junction device

In order to investigate the advance-delay effect involved in bubble production phenomenon in LCM processes, we have carried out experiments converging two immiscible fluids in a cylindrical T-shaped junction device (Fig. 2a) with inner radius of $R_c$ = 0.5mm and 1.0mm.

The injections are controlled by two syringe pumps and we name the flow rates of the continuous phase (either silicone oils or a glycerol/water mixture) and the dispersed phase (gaseous phase) respectively $Q_1$ and $Q_2$ (Fig. 2a). This simple device allows to create a steady train of calibrated bubbles by controlling both injection flow rates. An example of a confined breakup mechanism is showed in Fig. 1b for silicone oil 47V100 with a relatively low cross flow rate about $0.83 \mu L.s^{-1}$. A bubble emerges at the T-junction and grows as far as the opposite wall (Fig. 2b.*i*). The bubble neck is then pinched (Fig. 2b.*ii*) and breaks up (Fig. 2b.*iii*). The created bubble is then carried away by the cross flow.

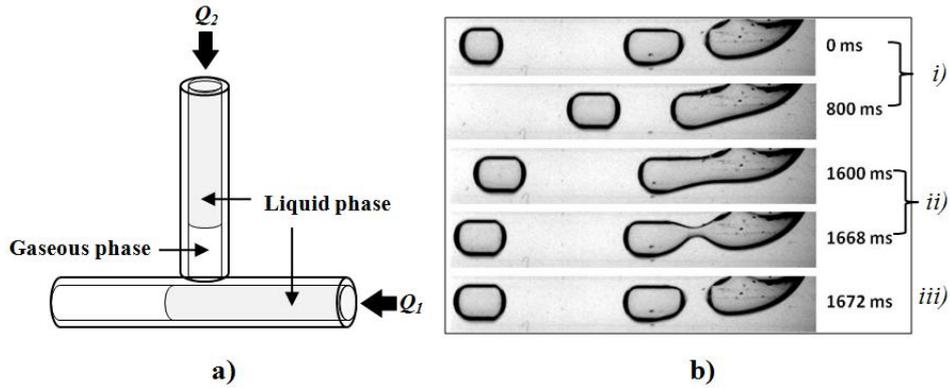

Fig. 2: (a) Sketch of the gas entrapped at the T-junction; (b) Bubble formation mechanism: *i)* filling phase, *ii)* squeezing phase and *iii)* break-up.

Garstecki et al. [9], who gave an empirical law for lengths of water droplets in oil, relate the coefficient $\alpha$ to the ratio of the both channel widths. In their approach, the authors use a coefficient $\beta$, which is equal to 1, meaning independent of flow conditions and the geometry. Recently, van Steijn et al. [11] made an extension to Garstecki's scaling rule in considering that the coefficient $\beta$ is linked to the volume of the bubble or the droplet at the filling time but yet independent from the flow conditions. They calculated both coefficients and concluded that they are fully characterized by the shape of the T-junction. These both analyses [9, 11] show that the bubble and droplet shapes are independent from the fluid properties.

For the squeezing regime, the experimental trends (Fig.3) show a normalized bubble length evolution depending linearly on the flow rate ratio with fitted coefficients $\alpha$ and $\beta$ (Eq. 1).

$$\frac{L}{2R_c} = \alpha \frac{Q_2}{Q_1} + \beta \qquad (1)$$

Note that measurements for both silicone oils are consistent with the linear model proposed by Garstecki et al. with $\alpha = 1$ (capillary radius ratio) and $\beta = 1$. However, for the glycerol/water mixture, experimental values are quite far from the Garstecki's model. Indeed, we obtain coefficients almost equal and close to $\alpha \approx \beta \approx 3$. Consequently, the coefficients ($\alpha$ and $\beta$) obtained are higher than those predicted by Garstecki et al. So the particular case where $\alpha = \beta = 1$ is also a relevant approximation for wetting liquids [9].

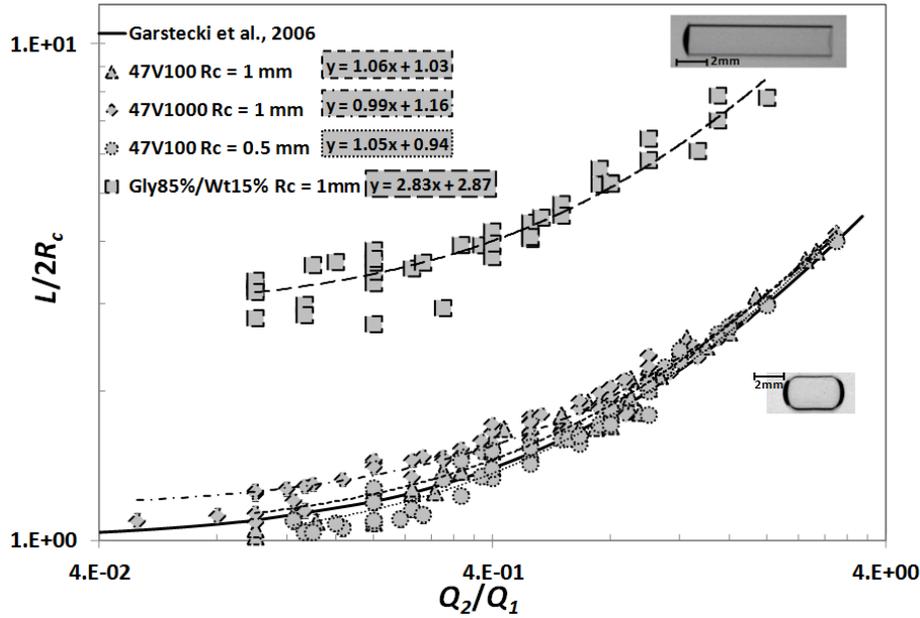

Fig. 3: Normalized bubble length as a function of flow rate ratio for the squeezing regime. Experimental fit: short dashed line for 47V100 ($R_c$ = 1mm); dotted line for 47V100 ($R_c$ = 0.5mm); dash dotted line for 47V1000 ($R_c$ = 1mm) and large dashed line for glycerol 85% - water 15% ($R_c$ = 1mm).

**Interconnected capillaries: supplying principle**

The pore doublet model studied in this part is composed of two circular capillaries with different radii: $R_M$, for the larger capillary, called "macrochannel" and $R_m$ for the smaller one, named "microchannel". The PDM capillaries are divided into two parts over the length: a first part where the capillaries are continuously interconnected on over a distance $l$, so-called "continuous interconnectivity", subsequently followed by a second part of length $L$, wherein the capillaries are interconnected at both ends forming nodes, so-named "node interconnectivity" (see Fig. 3a). Note that the void entrapped in the macrochannel and in the microchannel will be respectively called macrovoid (see Fig. 3b) and microvoid (see Fig. 3c). The filling of the PDM can be divided into two phases according to the position $x(t)$ of the both menisci into the two parts. The first phase, which is defined for the time range $\epsilon[0, t_{fl}]$, occurs when one of the both menisci reaches first the end of the first part ($x(t_{fl}) = l$). The second phase, defined for $t\epsilon[t_{fl}, t_{fL}]$, is when one of the both streams completely fills the second part ($x(t_{fL}) = l + L$). In order to simplify the following analytical development, the void is assumed to be incompressible and the liquid as Newtonian. Furthermore, we will neglect the gravity and the inertial effects and consider that the advanced contact angle at the intersection between the liquid-gas interface and the solid surface is supposed to be approximately equal to the apparent equilibrium contact angle (wetting liquids). To lighten the notations, we will use the subscripts $m$ and $M$, respectively for the microchannel and the macrochannel and the continuous interconnectivity and the node interconnectivity parts are noted with respectively the exponents $ci$ and $ni$. The governing flow equations describing the forces filling of the PDM is given below.

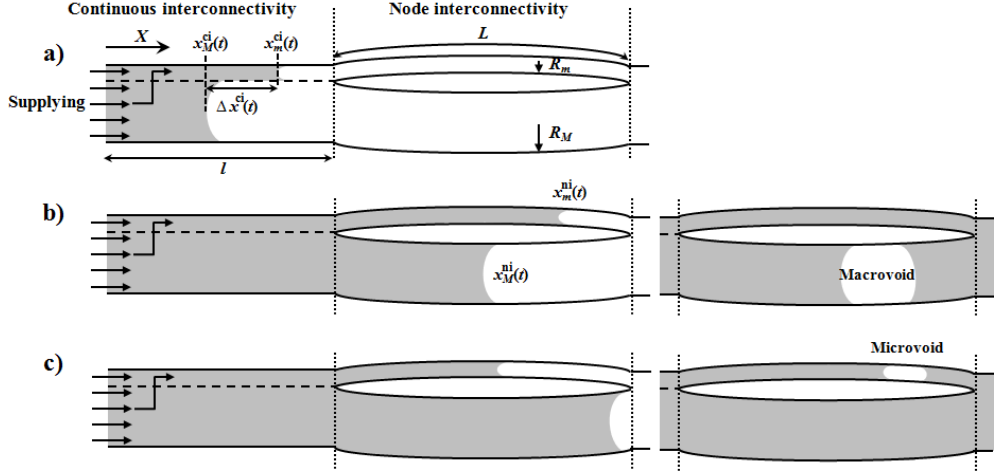

Fig.4: Sketch of void formation inside a PDM formed by a continuous interconnectivity and a node interconnectivity parts: a) Supplying principle, b) macrovoid and c) microvoid creations.

We consider the supplying principle [12] that consists in supplying mass to the microchannel from the macrochannel which plays the role of a tank. Consequently, the meniscus inside the microchannel is always ahead. Then, the difference between the both menisci $\Delta x^{ci}(t)$, given by balancing the viscous pressure drop with the capillary pressure between the both menisci, is as follows:

$$\Delta x(t) = \left(\frac{(R_M - R_m)^2 R_m^2}{R_M^3} \frac{\gamma_L \cos\theta_e}{2\eta}\right)^{1/2} t^{1/2} \quad (2)$$

For the first part of the macrochannel and for the second part of the both capillaries, the motion equation is deduced from balancing the pressures, with the addition of an injection pressure $P_i$ to the Lucas-Washburn equation [13, 14]:

$$\frac{8\eta}{R_{m,M}^2} x\dot{x} = P_i + \frac{2\gamma_L \cos\theta_e}{R_{m,M}} \quad (3)$$

Besides, we define a pressure $P^*$, for which the both menisci arrive at the same time $t_{fl}$ at the node $x = l$, expressed as follows:

$$P^* = \frac{2\gamma_L \cos\theta_e}{R_M} \left[\frac{\alpha(2 + \alpha - \alpha^2)}{1 + \alpha}\right] \quad (4)$$

Where $\alpha \epsilon [0,1]$ representing the ratio of capillary radii $R_m/R_M$. In the following, we set $\beta$ the part length ratio $L/l$.

- For $P_i < P^*$, we assume that the meniscus in the small capillary reaches first the node of the first part $x_m^{ci}(t_{fl}) = l$. From this configuration, two possibilities can occur: either *i)* the meniscus inside the macrochannel reaches first the second node $x_M^{ni}(t_{fL}) = l + L$, creating therefore the microvoid; or *ii)* the stream inside the microchannel remains in advance and reaches first the second node $x_m^{ni}(t_{fL}) = l + L$, hence the macrovoid is created.
    i. the difference between the both fronts at $t_{fL}$ can be expressed as follows:

$$\Delta x_m^{ni}(t_{fL}) = l\left(\beta - \alpha\left(\frac{P_iR_m+2\gamma_L\cos\theta_e}{P_iR_m+2\alpha\gamma_L\cos\theta_e}\right)^{1/2}\left(1+\beta - \frac{x_M^{ci}(t_{fl})}{l}\right)\right) \quad (5)$$

ii. the difference between the both menisci at $t_{fL}$ is given by:

$$\Delta x_M^{ni}(t_{fL}) = l\left(1+\beta - \frac{\beta}{\alpha}\left(\frac{P_iR_m+2\alpha\gamma_L\cos\theta_e}{P_iR_m+2\gamma_L\cos\theta_e}\right)^{1/2} - \frac{x_M^{ci}(t_{fl})}{l}\right) \quad (6)$$

- For $P_i > P^*$, we suppose that the flow inside the macrochannel reaches first the node at $x_M^{ci}(t_{fl}) = l$. In this configuration, we have only one possibility because in the one hand, the flow in the macrochannel fills first the first part and in the other hand, as the flow in the largest capillary is the fastest of the both, the stream in the small capillary couldn't catch up the other. Thus, the meniscus in the macrochannel will reach first the node at $x_M^{ni}(t_{fL}) = l + L$, so the microvoid is formed and the difference between the both front positions at $t_{fL}$ is given by:

$$\Delta x_m^{ni}(t_{fL}) = l\left(1+\beta - \alpha\beta\left(\frac{P_iR_m+2\gamma_L\cos\theta_e}{P_iR_m+2\alpha\gamma_L\cos\theta_e}\right)^{1/2} - \frac{x_m^{ci}(t_{fl})}{l}\right) \quad (7)$$

Curves plotted in Fig. 5 are given for $\alpha = 0.1$ and by varying the parameter $\beta$. We observe that for each value of $\beta$ and for $P_i < P^*$, the macrovoid rate decreases with the increase of the pressure $P_i$ because the stream inside the macrochannel is flowing increasingly quickly. Besides, for a given $P_i$, the macrovoid rate grows with the decrease of the parameter $\beta$. Moreover, one observes that when $P_i > P^*$, the microvoid rate is almost independent from $\beta$ and increases with $P_i$ until a limit value (about 1\%) corresponding to the volume rate of the microchannel in whole of the PDM.

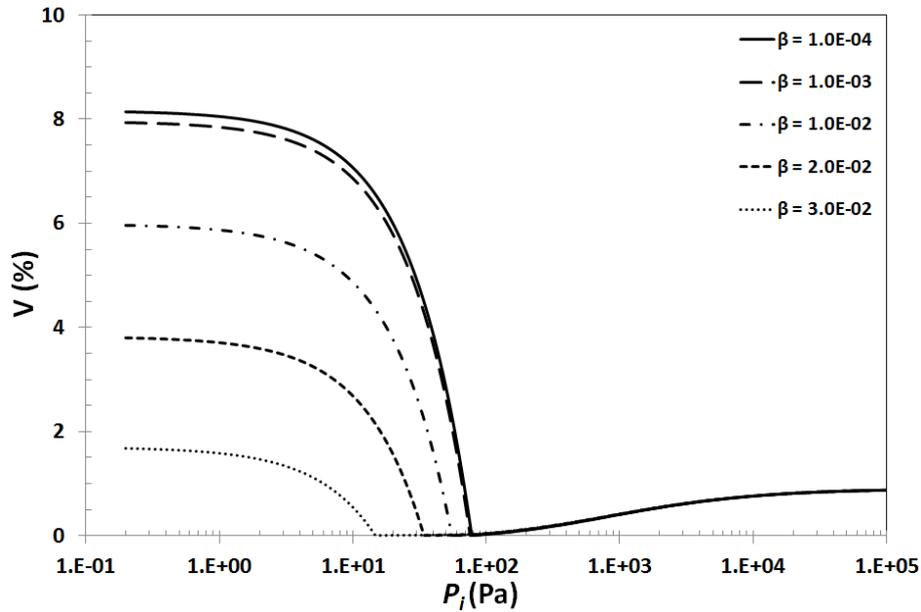

Fig. 5: Void rate evolution as a function of the imposed pressure $P_i$ for different values of $\beta$ with $\alpha = 0.1$, $l = 500$mm and $R_m = 10$μm.

Curves presented in Fig. 6 are obtained for $\beta = 2 \times 10^{-3}$ and different values of $\alpha$. First, at a given $\alpha$, the macrovoid rate decreases when $P_i$ increases for $P_i < P^*$ and the microvoid rate increases with $P_i$ when $P_i > P^*$. Secondly, at constant $P_i$, the void rate grows when $\alpha$ raises. This can be explained by the reduction of the total volume of the PDM when $\alpha$ increases (decrease of $R_M$). However, the values of $P_i$ relatively to $P^*$, influence more or less significantly this evolution. Actually, in the case of $P_i < P^*$, the decrease of the macrovoid with $\alpha$ is also notably due to a faster flow inside the macrochannel, reducing therefore the difference between the both meniscus positions at the end of the PDM. Moreover, the microvoid rate evolution is negligible compared to the macrovoid one. For $P_i > P^*$, the raise of the microvoid rate with $\alpha$ is mainly due to the decrease of the whole volume of the PDM.

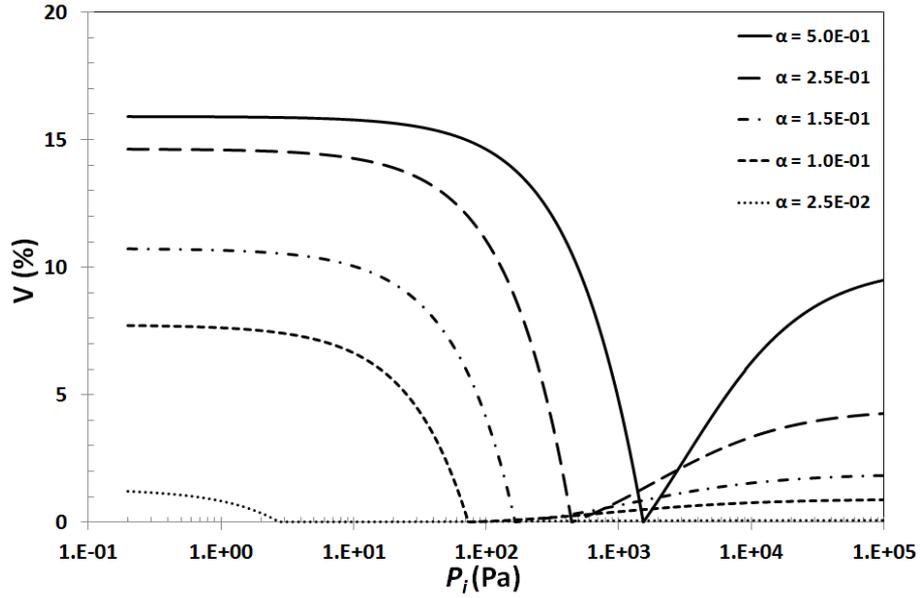

Fig. 6: Void rate evolution as a function of the imposed pressure for different values of $\alpha$ with $\beta = 0.002$, $l = 500$mm and $R_m = 10\mu$m.

## CONCLUSIONS

To conclude, in this paper, we have proposed issues to quantify the void created and entrapped in LCM processes. The measurements performed in the T-junction device emphasize the behaviour of non-wetting liquids (glycerol solution) which is quite different from wetting liquids (silicone oils). We showed that the Garstecki et al.'s model was well validated for wetting liquids but there is a discrepancy for the non-wetting liquid which has a higher value of the static contact angle. For instance, the chemical characterization of the polyester, vinylester and epoxy resins, that are partially wetting liquids, is a significant way to enhance LCM processes by improving the knowledge of the adhesion fibre/resin and notably the quantification of the void rate entrapped. Furthermore, we have proposed an analytic approach of void prediction based on both the Lucas-Washburn equation and the supplying principle for the imbibition case through an original PDM combining continuous and node interconnectivities.